\documentclass[runningheads]{llncs}
\usepackage{graphicx}
\usepackage{amsmath}
\usepackage{graphicx}
\usepackage{url}
\usepackage{dblfloatfix}
\usepackage{latexsym}
\usepackage{amssymb}
\usepackage{xcolor}

\usepackage{caption}

\begin{document}
\title{Embedding Geographic Locations for Modelling the Natural Environment using Flickr Tags and Structured Data}
\titlerunning{Embedding Geographic Locations (EGEL)}
%
\author{Shelan S. Jeawak \and Christopher B. Jones \and Steven Schockaert}

%
\authorrunning{S. Jeawak et al.}
%
\institute{Cardiff University, School of Computer Science and Informatics, Cardiff, UK
\email{\{JeawakSS,JonesCB2,SchockaertS1\}@cardiff.ac.uk}}
\maketitle              
\begin{abstract}
Meta-data from photo-sharing websites such as Flickr can be used to obtain rich bag-of-words descriptions of geographic locations, which have proven valuable, among others, for modelling and predicting ecological features. One important insight from previous work is that the descriptions obtained from Flickr tend to be complementary to the structured information that is available from traditional scientific resources. To better integrate these two diverse sources of information, in this paper we consider a method for learning vector space embeddings of geographic locations. We show experimentally that this method improves on existing approaches, especially in cases where structured information is available.

\keywords{Social media  \and Text mining \and Vector space embeddings \and Volunteered Geographic Information \and Biodiversity \and Ecology.}
\end{abstract}
\section{Introduction}
Users of photo-sharing websites such as Flickr\footnote{\url{http://www.flickr.com}} often provide short textual descriptions in the form of tags to help others find the images. With the availability of GPS systems in current electronic devices such as smartphones, latitude and longitude coordinates are nowadays commonly made available as well. The tags associated with such georeferenced photos often describe the location where these photos were taken, and Flickr can thus be regarded as a source of environmental information. The use of Flickr for modelling urban environments has already received considerable attention. For instance,  various approaches have  been proposed for modelling urban regions \cite{cunha2014using}, and for identifying points-of-interest \cite{DBLP:journals/ijswis/CanneytSD13} and itineraries \cite{de2010constructing,quercia2014shortest}. However, the usefulness of Flickr for characterizing the natural environment, which is the focus of this paper, is less well-understood.

Many recent studies have highlighted that Flickr tags capture valuable ecological information, which can be used as a complementary source to more traditional sources. To date, however, ecologists have mostly used social media to conduct manual evaluations of image content with little automated exploitation of the associated tags \cite{richards2015rapid,daume2016mining,ElQadi:2017}. One recent exception is \cite{jeawak2017using}, where bag-of-words representations derived from Flickr tags were found to give promising result for predicting a range of different environemental phenomena. 

Our main hypothesis in this paper is that by using vector space embeddings instead of bag-of-words representations, the ecological information which is implicitly captured by Flickr tags can be utilized in a more effective way. Vector space embeddings are representations in which the objects from a given domain are encoded using relatively low-dimensional vectors. They have proven useful in natural language processing, especially for encoding word meaning \cite{mikolov2013distributed,pennington2014glove}, and in machine learning more generally. In this paper, we are interested in the use of such representations for modelling geographic locations. Our main motivation for using vector space embeddings is that they allow us to integrate the textual information we get from Flickr with available structured information in a very natural way. To this end, we rely on an adaptation of the GloVe word embedding model \cite{pennington2014glove}, but rather than learning word vectors, we learn vectors representing locations. Similar to how the representation of a word in GloVe is determined by the context words surrounding it, the representation of a location in our model is determined by the tags of the photos that have been taken near that location. To incorporate numerical features from structured environmental datasets (e.g.\ average temperature), we associate with each such feature a linear mapping that can be used to predict that feature from a given location vector. This is inspired by the fact that salient properties of a given domain can often be modelled as directions in vector space embeddings \cite{gupta2015distributional,derracAIJ,DBLP:conf/acl/RotheS16}. Finally, evidence from categorical datasets (e.g.\ land cover types) is taken into account by requiring that locations belonging to the same category are represented using similar vectors, similar to how semantic types are sometimes modelled in the context of knowledge graph embedding \cite{guo2015semantically}. 

While our point-of-departure is a standard word embedding model, we found that the off-the-shelf GloVe model performed surprisingly poorly, meaning that a number of modifications are needed to achieve good results. Our main findings are as follows. First, given that the number of tags associated with a given location can be quite small, it is important to apply some kind of spatial smoothing, i.e.\ the importance of a given tag for a given location should not only depend on the occurrences of the tag at that location, but also on its occurrences at nearby locations. To this end, we use a formulation which is based on spatially smoothed version of pointwise mutual information. Second, given the wide diversity in the kind of information that is covered by Flickr tags, we find that term selection is in some cases critical to obtain vector spaces that capture the relevant aspects of geographic locations. For instance, many tags on Flickr refer to photography related terms, which we would normally not want to affect the vector representation of a given location\footnote{One exception is perhaps when we want to predict the scenicness of a given location, where e.g.\ terms that are related to professional landscape photography might be a strong indicator of scenicness.}. Finally, even with these modifications, vector space embeddings learned from Flickr tags alone are sometimes outperformed by bag-of-words representations. However, our vector space embeddings lead to substantially better predictions in cases where structured (scientific) information is also taken into account. In this sense, the main value of using vector space embeddings in this context is not so much about abstracting away from specific tag usages, but rather about the fact that such representations allow us to integrate numerical and categorical features in a much more natural way than is possible with bag-of-words representations.

The remainder of this paper is organized as follows. In the next section, we provide a discussion of existing work. Section \ref{region_embeddings} then presents our model for embedding geographic locations from Flickr tags and structured data. Next, in Section \ref{eval} we provide a detailed discussion about the experimental results. Finally, Section \ref{conclusion} summarizes our conclusions.
\section{Related Work}
\subsection{Vector space embeddings} The use of low-dimensional vector space embeddings for representing objects has already proven effective in a large number of applications, including natural language processing (NLP), image processing, and pattern recognition. In the context of NLP, the most prominent example is that of word embeddings, which represent word meaning using vectors of typically around 300 dimensions. A large number of different methods for learning such word embeddings have already been proposed, including Skip-gram and the Continuous Bag-of-Words (CBOW) model \cite{mikolov2013distributed}, GloVe \cite{pennington2014glove}, and fastText \cite{grave2017bag}. They have been applied effectively in many downstream NLP tasks such as sentiment analysis \cite{tang2014learning}, part of speech tagging \cite{qiu2014learning,liu2016part}, and text classification \cite{lilleberg2015support,ge2017improving}. The model we consider in this paper builds on GloVe, which was designed to capture linear regularities of word-word co-occurrence. In GloVe, there are two word vectors $w_i$ and $\tilde w_j$ for each word in the vocabulary, which are learned by minimizing the following objective:
\begin{equation*}\label{glove}
J=\sum_{i,j=1}^V{f(x_{ij})}{(w_i.\tilde w_j+b_i+\tilde b_j-\log x_{ij})^2}
\end{equation*}
where $x_{ij}$ is the number of times that word $i$ appears in the context of word $j$, $V$ is the vocabulary size, $b_i$ is the target word bias, $\tilde b_j$ is the context word bias. The weighting function $f$ is used to limit the impact of rare terms. It is defined as 1 if $x > x_{max}$ and as $(\frac {x}{x_{max}})^\alpha$ otherwise, where $x_{max}$ is usually fixed to 100 and $\alpha$ to 0.75. Intuitively, the target word vectors $w_i$ correspond to the actual word representations which we would like to find, while the context word vectors $\tilde{w_j}$ model how occurrences of $j$ in the context of a given word $i$ affect the representation of this latter word. In this paper we will use a similar model, which will however be aimed at learning location vectors instead of the target word vectors.

Beyond word embeddings, various methods have been proposed for learning vector space representations from structured data such as knowledge graphs \cite{NIPS20135071,yang2014embedding,ComplEx}, social networks \cite{grover2016node2vec,wang2017community} and taxonomies \cite{vendrov2015order,DBLP:conf/nips/NickelK17}. The idea of combining a word embedding model with structured information has also been explored by several authors, for example to improve the word embeddings based on information coming from knowledge graphs \cite{DBLP:conf/cikm/XuBBGWLL14,speer2017conceptnet}. Along similar lines, various lexicons have been used to obtain word embeddings that are better suited at modelling sentiment \cite{tang2014learning} and antonymy \cite{ono2015word}, among others. The method proposed by \cite{liu2015learning} imposes the condition that words that belong to the same semantic category are closer together than words from different categories, which is somewhat similar in spirit to how we will model categorical datasets in our model.

\subsection{Embeddings for geographic information}
The problem of representing geographic locations using embeddings has also attracted some attention. An early example is
\cite{saeidi2015lower}, which used principal component analysis and stacked autoencoders to learn low-dimensional vector representations of city neighbourhoods based on census data. They use these representations to predict attributes such as crime, which is not included in the given census data, and find that in most of the considered evaluation tasks, the low-dimensional vector representations lead to more faithful predictions than the original high-dimensional census data. 

Some existing works combine word embedding models with geographic coordinates. For example, in \cite{cocos2017language} an approach is proposed to learn word embeddings based on the assumption that words which tend to be used in the same geographic locations are likely to be similar. Note that their aim is dual to our aim in this paper: while they use geographic location to learn word vectors, we use textual descriptions to learn vectors representing geographic locations. 

Several methods also use word embedding models to learn representations of Points-of-Interest (POIs) that can be used for predicting user visits \cite{FengCAC17,Liu:-IJCAI-2016,SZhaoWWW:2017}. These works use the machinery of existing word embedding models to learn POI representations, intuitively by letting sequences of POI visits by a user play the role of sequences of words in a sentence. In other words, despite the use of word embedding models, many of these approaches do not actually consider any textual information. For example, in \cite{Liu:-IJCAI-2016} the Skip-gram model is utilized to create a global pattern of users' POIs. Each location was treated as a word and the other locations visited before or after were treated as context words. They then use a pair-wise ranking loss \cite{Weston:2010} which takes into account the user's location visit frequency to personalize the location recommendations. The methods of \cite{Liu:-IJCAI-2016} were extended in \cite{SZhaoWWW:2017} to use a temporal embedding and to take more account of geographic context, in particular the distances between preferred and non-preferred neighboring POIs, to create a ``geographically hierarchical pairwise preference ranking model''. Similarly, in \cite{Yao-IJGIS:2017} the CBOW model was trained with POI data. They ordered POIs spatially within the traffic-based zones of urban areas. The ordering was used to generate characteristic vectors of POI types. Zone vectors represented by averaging the vectors of the POIs contained in them, were then used as features to predict land use types. In the CrossMap method \cite{CZhangCrossModal-WWW:2017} they learned embeddings for spatio-temporal hotspots obtained from social media data of locations, times and text. In one form of embedding, intended to enable reconstruction of records, neighbourhood relations in space and time were encoded by averaging  hotspots in a target location's spatial and temporal neighborhoods. They also proposed a graph-based embedding method with nodes of location, time and text. The concatenation of the location, time and text vectors were then used as features to predict peoples' activities in urban environments.  Finally, in \cite{Yan-Janowicz:2017}, a method is proposed that uses the Skip-gram model to represent POI types, based on the intuition that the vector representing a given POI type should be predictive of the POI types that found near places of that type.

Our work is different from these studies, as our focus is on representing locations based on a given text description of that location (in the form of Flickr tags), along with numerical and categorical features from scientific datasets.



\subsection{Analyzing Flickr tags}
Many studies have focused on analyzing Flickr tags to extract useful information in domains such as linguistics \cite{eisenstein2010latent}, geography \cite{cunha2014using,grothe2009automated}, and ecology \cite{barve2015discovering,jeawak2017using,jeawak2018mapping}. Most closely related to our work, \cite{jeawak2017using} found that the tags of georeferenced Flickr photos can effectively supplement traditional scientific environmental data in tasks such as predicting climate features, land cover, species occurrence, and human assessments of scenicness. To encode locations, they simply combine a bag-of-words representation of geographically nearby tags with a feature vector that encodes associated structured scientific data. They found that the predictive value of Flickr tags is roughly on a par with that of the scientific datasets, and that combining both types of information leads to significantly better results than using either of them alone. As we show in this paper, however, their straightforward way of combining both information sources, by concatenating the two types of feature vectors, is far from optimal. 

Despite the proven importance of Flickr tags, the problem of embedding Flickr tags has so far received very limited attention. To the best of our knowledge, \cite{hasegawa2018social} is the only work that generated embeddings for Flickr tags. However, their focus was on learning embeddings that capture word meaning (being evaluated on word similarity tasks), whereas we use such embeddings as part of our method for representing locations.


\section{Model Description} \label{region_embeddings}
In this section, we introduce our embedding model, which combines Flickr tags and structured scientific information to represent a set of  locations $L$. The proposed model has the following form:
\begin{equation}\label{Region_model}
J=\alpha J_{\textit{tags}}+(1-\alpha)J_{\textit{nf}}+\beta J_{\textit{cat}}
\end{equation}
where $\alpha \in [0, 1]$ and $\beta \in [0, +\infty]$ are parameters to control the importance of each component in the model. Component $J_{\textit{tags}}$ will be used to constrain the representation of the locations based on their textual description (i.e.\ Flickr tags), $J_{\textit{nf}}$ will be used to constrain the representation of the locations based on their numerical features, and $J_{\textit{cat}}$ will impose the constraint that locations belonging to the same category should be close together in the space. We will discuss each of these components in more detail in the following sections.

\subsection{Tag Based Location Embedding} \label{tags_embeddings}
Many of the tags associated with Flickr photos describe characteristics of the places where these photos were taken \cite{hollenstein2010exploring,rattenbury2007towards,rattenbury2009methods}. For example, tags may correspond to place names (e.g.\ Brussels, England, Scandinavia), landmarks (e.g.\ Eiffel Tower, Empire State Building) or land cover types (e.g.\ mountain, forest, beach). To allow us to build location models using such tags, we collected the tags and meta-data of 70 million Flickr photos with coordinates in Europe (which is the region our experiments will focus on), all of which were uploaded to Flickr before the end of September 2015. In this section we first explain how tags can be weighted to obtain bag-of-words representations of locations from Flickr. Subsequently we describe a tag selection method, which will allow us to specialize the embedding depending on which aspects of the considered locations are of interest, after which we discuss the actual embedding model.

\smallskip
\noindent \textbf{Tag weighting.} Let $L=\{l_1,...,l_m\}$ be a set of geographic locations, each characterized by latitude and longitude coordinates. 
To generate a bag-of-words representation of a given location, we have to weight the relevance of each tag to that location. To this end, we have followed the weighting scheme from \cite{jeawak2017using}, which combines a Gaussian kernel (to model spatial proximity) with Positive Pointwise Mutual Information (PPMI) \cite{church1990word,niwa1994co}.

Let us write $U_{t,l}$ for the set of users who have assigned tag $t$ to a photo with coordinates near $l$. To assess how relevant $t$ is to the location $l$, the number of times $t$ occurs in photos near $l$ is clearly an important criterion. However, rather than simply counting the number of occurrences within some fixed radius, we use a Gaussian kernel to weight the tag occurrences according to their distance from that location:
\begin{equation*}
w(t,l)=\sum_{d(l,r)\leq D} |U_{t,l}|\cdot \exp\Big(-\frac{d^2\big(l,r\big)}{2\sigma^2}\Big)
\end{equation*}
where the threshold $D>0$ is assumed to be fixed, $r$ is the location of a Flickr photo, $d$ is the Haversine distance, and we will assume that the bandwidth parameter $\sigma$ is set to $D/3$. A tag occurrence is counted only once for all photos by the same user at the same location, which is important to reduce the impact of bulk uploading. The value $w(t,l)$ reflects how frequent tag $t$ is near location $l$, but it does not yet take into account the total number of tag occurrences near $l$, nor how popular the tag $t$ is overall. To measure how strongly tag $t$ is associated with location $l$, we use PPMI, which is a commonly used measure of association in natural language processing. However, rather than estimating PPMI scores from term frequencies, we will use the $w(t,l)$ values instead:
    \begin{equation*}
    \textit{PPMI}(t,l)=\max\left(0,\log\left(\frac{p_{t,l}}{p_tp_{l}}\right)\right)
    \end{equation*}
    where:
    \begin{align*}
    p_{t,l}&=\frac{w(t,l)}{N} 
    &
    p_t&= \frac{\sum_{l'\in L}{w(t,l')}}{N}
    &
    N&=\sum_{t'\in T}\sum_{l'\in L}w(t',l')
    &
    p_{l}&=\frac{\sum_{t'\in T}{w(t',l)}}{N}
    \end{align*}
with $T$ the set of all tags, and $L$ the set of locations.

\smallskip
\noindent \textbf{Tag selection.}
Inspired by \cite{kuang2017learning}, we use a term selection method in order to focus on the tags that are most important for the tasks that we want to consider and reduce the impact of tags that might relate only to a given individual or a group of users. 
In particular, we obtained good results with a method
based on Kullback-Leibler (KL) divergence, which is based on \cite{DBLP:journals/tkde/LaereQSD14}. 
Let $C_1,...,C_n$ be a set of (mutually exclusive) properties of locations in which we are interested (e.g.\ land cover categories). For the ease of presentation, we will identify $C_i$ with the set of locations that have the corresponding property. Then, we select tags from $T$ that maximize the following score:
\begin{equation*}
\textit{KL}(t)=\sum_{i=1}^n{P(C_i|t)}\log\frac {P(C_i|t)}{Q(C_i)}
\end{equation*}
where $P(C_i|t)$ is the probability that a photo with tag $t$ has a location near $C_i$ and $Q(C_i)$ is the probability that an arbitrary tag occurrence is assigned to a photo near a location in $C_i$. Since $P(C_i|t)$ often has to be estimated from a small number of tag occurrences, it is estimated using Bayesian smoothing:
\begin{equation*}
P(C_i | t)=\frac {\left(\sum_{l\in C_i} w(t,l)\right)  + \gamma \cdot Q(C_i)}{N+\gamma}
\end{equation*}
where $\gamma$ is a parameter controlling the amount of smoothing, which will be tuned in the experiments. On the other hand, for $Q(C_i)$ we can simply use a maximum likelihood estimation:
\begin{align*}
Q(C_i)&=\frac{\sum_{l\in C_i} \sum_{t\in T}{w(t,l)}}{\sum_{j=1}^n \sum_{l\in C_j} \sum_{t\in T} w(t,l)}
\end{align*}
\smallskip
\noindent \textbf{Location embedding.}
We now want to find a vector $v_{l_i} \in V$ for each location $l_i$ such that similar locations are represented using similar vectors. To achieve this, we use a close variant of the GloVe model, where tag occurrences are treated as context words of geographic locations. In particular, with each location $l$ we associate a vector $v_l$ and with each tag $t$ we associate a vector $\tilde{w_t}$ and a bias term $\tilde{b_{t_j}}$, and consider the following objective (which in our full model \eqref{Region_model} will be combined  with components that are derived from the structured information):
\begin{align*}
J_{\textit{tags}}=\sum_{l_i \in L}\sum_{t_j \in T}(v_{l_i} \tilde{w_{t_j}} + \tilde{b_{t_j}} - \textit{PPMI}(t_j,l_i))^2
\end{align*}
Note how tags play the role of the context words in the GloVe model, while instead of learning target word vectors we now learn location vectors. In contrast to GloVe, our objective does not directly refer to co-occurrence statistics, but instead uses the $\textit{PPMI}(t_j,l_i)$ scores. One important consequence of this is that we can also consider pairs $(l_i,t_j)$ for which $t_j$ does not occur in $l_i$ at all; such pairs are usually called \emph{negative examples}. While they cannot be used in the standard GloVe model, some authors have already reported that introducing negative examples in variants of GloVe can lead to an improvement \cite{jameel2016d}. 
In practice, evaluating the full objective above would not be computationally feasible, as we may need to consider millions of locations and millions of tags. Therefore, rather than considering all tags in $T$ for the inner summation, we only consider those tags that appear at least once near location $l_i$ together with a sample of negative examples.
\subsection{Structured Environmental Data} \label{struc_embeddings}
There is a wide variety of structured data that can be used to describe locations. In this work, we have restricted ourselves to the same datasets as \cite{jeawak2017using}. These include nine (real-valued) numerical features, which are latitude, longitude, elevation\footnote{\url{http://www.eea.europa.eu/data-and-maps/data/eu-dem}}, population\footnote{\url{http://data.europa.eu/89h/jrc-luisa-europopmap06}}, and five climate\footnote{\url{http://worldclim.org}} related features (avg.\ temperature, avg.\ precipitation, avg.\ solar radiation, avg.\ wind speed, and avg.\ water vapor pressure). In addition, 180 categorical features were used,  which are CORINE\footnote{\url{http://www.eea.europa.eu/data-and-maps/data/corine-land-cover-2006-raster-2}} land cover classes at level 1 (5 classes), level 2 (15 classes) and level 3 (44 classes) and 116 soil types (SoilGrids\footnote{\url{https://www.soilgrids.org}}). Note that each location should belong to exactly 4 categories: one CORINE class at each of the three levels and a soil type. 

\smallskip
\noindent \textbf{Numerical features.}
Numerical features can be treated similarly to the tag occurrences, i.e.\ we will assume that the value of a given numerical feature can be predicted from the location vectors using a linear mapping. In particular, for each numerical feature $f_k$ we consider a vector $\tilde{w_{f_k}}$ and a bias term $\tilde{b_{f_k}}$, and the following objective:
\begin{align*}
J_{\textit{nf}}=\sum_{l_i \in L}\sum_{f_k \in \textit{NF}}(v_{l_i}. \tilde{w_{f_k}} + \tilde{b_{f_k}} - \textit{score}(f_k,l_i))^2
\end{align*}
where we write $\textit{NF}$ for set of all numerical features and $\textit{score}(f_k,l_i)$ is the value of feature $f_k$ for location $l_i$, after z-score normalization.

\smallskip
\noindent \textbf{Categorical features.}
To take into account the categorical features, we impose the constraint that locations belonging to the same category should be close together in the space. To formalize this, we represent each category type $\textit{cat}_l$ as a vector $w_{\textit{cat}_l}$, and consider the following objective:
\begin{align*}
J_{\textit{cat}}=\sum_{l_i \in R}{\sum_{cat_l \in C}(v_{l_i}- w_{cat_l})^2}
\end{align*}

\section{Experimental Results}\label{eval}
\subsection{Evaluation Tasks}
We will use the method from \cite{jeawak2017using} as our main baseline. This will allow us to directly evaluate the effectiveness of embeddings for the considered problem, since we have used the same structured datasets and same tag weighting scheme. For this reason, we will also follow their evaluation methodology. In particular, we will consider three evaluation tasks: 
   \begin{enumerate}
     \item Predicting the distribution of 100 species across Europe, using the European network of nature protected sites Natura 2000\footnote{\url{http://ec.europa.eu/environment/nature/natura2000/index_en.htm}} dataset as ground truth. For each of these species, a binary classification problem is considered. The set of locations $L$ is defined as the 26,425 distinct sites occurring in the dataset. 
     \item Predicting soil type, again each time treating the task as a binary classification problem, using the same set of locations $L$ as in the species distribution experiments. For these experiments, none of the soil type features are used for generating the embeddings.
     \item Predicting CORINE land cover classes at levels 1, 2 and level 3, each time treating the task as a binary classification problem, using the same set of locations $L$ as in the species distribution experiments. For these experiments, none of the CORINE features are used for generating the embeddings. 
   \end{enumerate}
In addition, we will also consider the following regression tasks:   
     \begin{enumerate}
     \item Predicting 5 climate related features: the average precipitation, temperature, solar radiation, water vapor pressure, and wind speed. We again use the same set of locations $L$ as for species distribution in this experiment. None of the climate features is used for constructing the embeddings for this experiment. 
     \item Predicting people's subjective opinions of landscape beauty in Britain, using the crowdsourced dataset from the ScenicOrNot website\footnote{\url{http://scenic.mysociety.org/}} as ground truth. The set $L$ is chosen as the set of locations of 191\,605 rated locations from the ScenicOrNot dataset for which at least one georeferenced Flickr photo exists within a 1 km radius. 
   \end{enumerate}
\begin{table}
  \begin{minipage}{0.52\textwidth}
  \begin{center} 
	\begin{footnotesize}
	\begin{tabular}{ |c|c| }
       \hline
            &Prec\hspace{0.35cm}Rec\hspace{0.35cm}F1\\
        \hline
            BOW-Tags &0.57\hspace{0.25cm}0.11\hspace{0.25cm}0.18\\
            BOW-KL(Tags)&0.11\hspace{0.25cm}0.86\hspace{0.25cm}0.19\\
            GloVe &0.10\hspace{0.25cm}0.88\hspace{0.25cm}0.17\\  
    	    EGEL-Tags  &0.10\hspace{0.25cm}0.88\hspace{0.25cm}0.18\\ 
    	    EGEL-Tags+NS &0.12\hspace{0.25cm}0.82\hspace{0.25cm}0.21\\ 
    	    EGEL-KL(Tags+NS) &0.15\hspace{0.25cm}0.64\hspace{0.25cm}0.25\\ 
            BOW-All&0.65\hspace{0.25cm}0.50\hspace{0.25cm}0.56\\
    	    EGEL-All &0.56\hspace{0.25cm}0.60\hspace{0.25cm}\textbf{0.58}\\
       \hline
  	\end{tabular}
  	\end{footnotesize}
  	\caption{Results for predicting species distribution.\label{tab_species}}
    \end{center}
    \end{minipage}
        \begin{minipage}{0.47\textwidth}
    \begin{footnotesize}
    \begin{tabular}{ |c|c|  }
       \hline
            &Prec\hspace{0.35cm}Rec\hspace{0.35cm}F1\\
        \hline
            BOW-Tags &0.17\hspace{0.25cm}0.44\hspace{0.25cm}0.24\\
            BOW-KL(Tags)&0.30\hspace{0.25cm}0.43\hspace{0.25cm}0.36\\
            GloVe &0.32\hspace{0.25cm}0.39\hspace{0.25cm}0.35\\  
    	    EGEL-Tags  &0.32\hspace{0.25cm}0.40\hspace{0.25cm}0.36\\ 
    	    EGEL-Tags+NS &0.30\hspace{0.25cm}0.44\hspace{0.25cm}0.36\\ 
    	    EGEL-KL(Tags+NS) &0.32\hspace{0.25cm}0.44\hspace{0.25cm}0.37\\ 
            BOW-All &0.39\hspace{0.25cm}0.43\hspace{0.25cm}0.41\\
    	    EGEL-All &0.33\hspace{0.25cm}0.67\hspace{0.25cm}\textbf{0.44}\\
       \hline
    \end{tabular}
  \end{footnotesize}
   \caption{Results for predicting soil type.\label{tab_soil}}
  \end{minipage}
\end{table}

\begin{table}
\centering
    \begin{footnotesize}
    \begin{tabular}{ |c|c|c|c|  }
       \hline
            &CORINE level 1 & CORINE level 2 & CORINE level 3\\
            &Prec\hspace{0.4cm}Rec\hspace{0.4cm}F1 &Prec\hspace{0.4cm}Rec\hspace{0.4cm}F1 & Prec\hspace{0.4cm}Rec\hspace{0.4cm}F1\\
        \hline
            BOW-Tags &0.49\hspace{0.3cm}0.45\hspace{0.3cm}0.47& 0.20\hspace{0.3cm}0.13\hspace{0.3cm}0.16& 0.14\hspace{0.3cm}0.08\hspace{0.3cm}0.10\\
            BOW-KL(Tags)&0.40\hspace{0.3cm}0.47\hspace{0.3cm}0.43&0.39\hspace{0.3cm}0.12\hspace{0.3cm}0.18&0.24\hspace{0.3cm}0.13\hspace{0.3cm}0.17\\
            GloVe &0.20\hspace{0.3cm}0.90\hspace{0.3cm}0.33&0.12\hspace{0.3cm}0.53\hspace{0.3cm}0.19&0.12\hspace{0.3cm}0.25\hspace{0.3cm}0.17\\  
    	    EGEL-Tags  &0.20\hspace{0.3cm}0.89\hspace{0.3cm}0.33&0.12\hspace{0.3cm}0.56\hspace{0.3cm}0.20&0.16\hspace{0.3cm}0.21\hspace{0.3cm}0.18\\ 
    	    EGEL-Tags+NS &0.23\hspace{0.3cm}0.73\hspace{0.3cm}0.35&0.12\hspace{0.3cm}0.52\hspace{0.3cm}0.20&0.18\hspace{0.3cm}0.22\hspace{0.3cm}0.19\\ 
    	    EGEL-KL(Tags+NS) &0.26\hspace{0.3cm}0.62\hspace{0.3cm}0.37&0.14\hspace{0.3cm}0.58\hspace{0.3cm}0.23&0.19\hspace{0.3cm}0.25\hspace{0.3cm}0.22\\ 
            BOW-All&0.52\hspace{0.3cm}0.51\hspace{0.3cm}0.51& 0.27\hspace{0.3cm}0.19\hspace{0.3cm}0.22& 0.18\hspace{0.3cm}0.11\hspace{0.3cm}0.13\\
    	    EGEL-All &0.45\hspace{0.3cm}0.66\hspace{0.3cm}\textbf{0.54}&0.27\hspace{0.3cm}0.48\hspace{0.3cm}\textbf{0.35}&0.23\hspace{0.3cm}0.33\hspace{0.3cm}\textbf{0.27}\\
       \hline
    \end{tabular}
  \end{footnotesize}
  \caption{Results for predicting CORINE land cover classes, at levels 1, 2 and 3.\label{tab_CORINE}}
\end{table}
\begin{table}
  \begin{center} 
	\begin{footnotesize}
	\begin{tabular}{ |c|c|c|c|c|c|  }
    	\hline
    	&Temp&Precip&Solar rad&Water vap&Wind speed\\
     	&MAE\hspace{0.5cm}$\rho$&MAE\hspace{0.5cm}$\rho$&MAE\hspace{0.5cm}$\rho$&MAE\hspace{0.5cm}$\rho$&MAE\hspace{0.5cm}$\rho$\\
    \hline
    	BOW-Tags  &1.62\hspace{0.3cm}0.84&11.66\hspace{0.3cm}0.68&926\hspace{0.4cm}0.83&0.08\hspace{0.3cm}0.71&0.54\hspace{0.3cm}0.75\\
    	BOW-KL(Tags) &1.69\hspace{0.3cm}0.81&12.85\hspace{0.3cm}0.65&1057\hspace{0.3cm}0.75&0.08\hspace{0.3cm}0.71&0.53\hspace{0.3cm}0.73\\
    	GloVe  &1.96\hspace{0.3cm}0.44&15.37\hspace{0.3cm}0.31&1507\hspace{0.3cm}0.36&0.11\hspace{0.3cm}0.47&0.74\hspace{0.3cm}0.28\\  
    	EGEL-Tags&1.95\hspace{0.3cm}0.47&15.03\hspace{0.3cm}0.31&1426\hspace{0.3cm}0.41&0.10\hspace{0.3cm}0.46&0.73\hspace{0.3cm}0.32\\ 
    	EGEL-Tags+NS&1.97\hspace{0.3cm}0.44&14.93\hspace{0.3cm}0.32&1330\hspace{0.3cm}0.44&0.10\hspace{0.3cm}0.46&0.72\hspace{0.3cm}0.36\\ 
    	EGEL-KL(Tags+NS)&1.48\hspace{0.3cm}0.73&13.55\hspace{0.3cm}0.52&1008\hspace{0.3cm}0.77&0.08\hspace{0.3cm}0.66&0.65\hspace{0.3cm}0.59\\ 
    	BOW-All &0.72\hspace{0.3cm}0.94&10.52\hspace{0.3cm}0.75&484\hspace{0.3cm}0.93&\textbf{0.05}\hspace{0.3cm}0.91&\textbf{0.43}\hspace{0.3cm}0.84\\ 
    	EGEL-All &\textbf{0.71}\hspace{0.3cm}\textbf{0.95}&\textbf{10.03}\hspace{0.3cm}\textbf{0.79}&\textbf{436}\hspace{0.3cm}\textbf{0.95}&\textbf{0.05}\hspace{0.3cm}\textbf{0.92}&\textbf{0.43}\hspace{0.3cm}\textbf{0.88}\\
    \hline
  	\end{tabular}
  	\end{footnotesize}
  	\caption{Results for predicting average climate data.\label{tab_climate} }
  \end{center}	
\end{table}
\subsection{Experimental Setup} 
In all experiments, we use Support Vector Machines (SVMs) for classification problems and Support Vector Regression (SVR) for regression problems to make predictions from our representations of geographic locations. In both cases, we used the SVM$^{\textit{light}}$ implementation\footnote{\url{http://www.cs.cornell.edu/people/tj/svm_light/}} \cite{joachims1998making}.  For each experiment, the set of locations $L$ was split into two-thirds for training, one-sixth for testing, and one-sixth for tuning the parameters. 
All embedding models are learned with Adagrad using 30 iterations. The number of dimensions is chosen for each experiment from $\{10,50,300\}$ based on the tuning data.
For the parameters of our model in Equation \ref{Region_model}, we considered values of $\alpha$ from \{0.1, 0.01, 0.001, 0.0001\} and values of $\beta$ from \{1, 10, 100, 1000, 10\,000, 100\,000\}. 

To compute KL divergence, we need to determine a set of classes $C_1,...,C_n$ for each experiment. For classification problems, we can simply consider the given categories, but for the regression problems we need to define such classes by discretizing the numerical values. For the scenicness experiments, we considered scores 3 and 7 as cut-off points, leading to three classes (i.e.\ less than 3, between 3 and 7, and above 7). Similarly, for each climate related features, we consider two cut-off values for discretization: 5 and 15 for average temperature, 50 and 100 for average precipitation, 10\,000 and 17\,000 for average solar radiation, 0.7 and 1 for average water vapor pressure, and 3 and 5 for wind speed. 
The smoothing parameter $\gamma$ was selected among $\{10, 100, 1000\}$ based on the tuning data. In all experiments where term selection is used, we select the top 100\,000 tags.  We fixed the radius $D$ at 1km when counting  the  number  of  tag occurrences. Finally, we set the number of negative examples as 10 times the number of positive examples for each location, but with a cap at 1000 negative examples in each region for computational reasons. We tune all parameters with respect to the F1 score for the classification tasks, and Spearman $\rho$ for the regression tasks.

\subsection{Variants and Baseline Methods}
We will refer to our model as EGEL (Embedding GEographic Locations), and will consider the following variants. 
\textbf{EGEL-Tags} only uses the information from the Flickr tags (i.e.\ component $J_{\textit{tags}}$), without using any negative examples and without feature selection.
\textbf{EGEL-Tags+NS} is similar to \textbf{EGEL-Tags} but with the addition of negative examples.
\textbf{EGEL-KL(Tags+NS)} additionally considers term selection.
\textbf{EGEL-All} is our full method, i.e.\ it additionally uses the structured information. We also consider the following baselines.
\textbf{BOW-Tags} represents locations using a bag-of-words representation, using the same tag weighting as the embedding model.
\textbf{BOW-KL(Tags)} uses the same representation but after term selection, using the same KL-based method as the embedding model.
\textbf{BOW-All} combines the bag-of-words representation with the structured information, encoded as proposed in \cite{jeawak2017using}. 
\textbf{GloVe} uses the objective from the original GloVe model for learning location vectors, i.e.\ this variant differs from \textbf{EGEL-Tags} in that instead of $\textit{PPMI}(t_j,l_i)$ we use the number of co-occurrences of tag $t_j$ near location $l_i$, measured as $|U_{t_jl_i}|$.

\begin{table}
\centering
\begin{minipage}{0.4\textwidth}
    \begin{footnotesize}
    \begin{tabular}{ |c|c| }
    \hline
     	& MAE\hspace{0.5cm}$\rho$\\
    \hline
    	BOW-Tags  &1.01\hspace{0.3cm}0.57\\
    	BOW-KL(Tags) &1.09\hspace{0.3cm}0.51\\
    	GloVe  &1.27\hspace{0.3cm}0.19\\
    	EGEL-Tags  &1.12\hspace{0.3cm}0.37\\
    	EGEL-Tags+NS &1.14\hspace{0.3cm}0.40\\
    	EGEL-KL(Tags+NS) &1.05\hspace{0.3cm}0.53\\
    	BOW-All &1.00\hspace{0.3cm}0.58\\
    	EGEL-All & \bf{0.94}\hspace{0.3cm}\bf{0.64}\\
    \hline
    \end{tabular}
  \end{footnotesize}
   \caption{Results for predicting scenicness.\label{tab_scenic}}
\end{minipage}\hfill
\begin{minipage}{0.53\textwidth}
\centering
	\includegraphics[width=170pt]{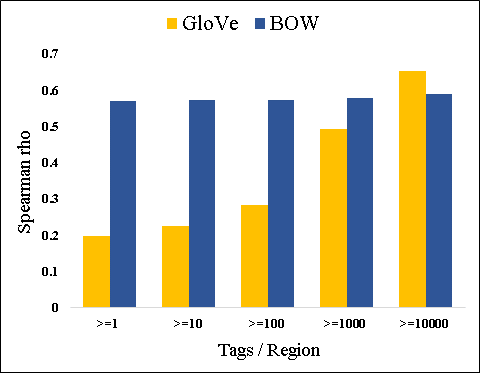}
	\captionof{figure}{Comparison between the performance of the GloVe and bag-of-words models for predicting scenicness, as a function of the number of tag occurrences at the considered locations.\label{fig_GloVe_BOW}}
\end{minipage}
\end{table}

\subsection{Results and Discussion}
We present our results for the binary classification tasks in Tables \ref{tab_species}--\ref{tab_CORINE} in terms of average precision, average recall and macro average F1 score. The results of the regression tasks are reported in Tables  \ref{tab_climate} and \ref{tab_scenic} in terms of the mean absolute error between the predicted and actual scores, as well as the Spearman $\rho$ correlation between the rankings induced by both sets of scores. It can be clearly seen from the results that our proposed method (EGEL-All) can effectively integrate Flickr tags with the available structured information. It outperforms the baselines for all the considered tasks. 
Furthermore, note that the PPMI-based weighting in EGEL-Tags consistently outperforms GloVe and that both the addition of negative examples and term selection lead to further improvements. The use of term selection leads to particularly substantial improvements for the regression problems.

While our experimental results confirm the usefulness of embeddings for predicting environmental features, this is only consistently the case for the variants that use both the tags and the structured datasets. In particular, comparing BOW-Tags with EGEL-Tags, we sometimes see that the former achieves the best results. While this might seem surprising, it is in accordance with the findings in \cite{joulin2017fast,CZhangCrossModal-WWW:2017}, among others, where it was also found that bag-of-words representations can sometimes lead to surprisingly effective baselines. Interestingly, we note that in all cases where EGEL-KL(Tags+NS) performs worse than BOW-Tags, we also find that BOW-KL(Tags) performs worse than BOW-Tags. This suggests that for these tasks there is a very large variation in the kind of tags that can inform the prediction model, possibly including e.g.\ user-specific tags. Some of the information captured by such highly specific but rare tags is likely to be lost in the embedding.

To further analyze the difference in performance between BoW representations and embeddings, Figure \ref{fig_GloVe_BOW} compares the performance of the GloVe model with the bag-of-words model for predicting place scenicness, as a function of the number of tag occurrences at the considered locations. What is clearly noticeable in Figure \ref{fig_GloVe_BOW} is that GloVe performs better than the bag-of-words model for large corpora and worse for smaller corpora.  This issue has been alleviated in our embedding method by the addition of negative examples. 

\section{ Conclusions}\label{conclusion}
In this paper, we have proposed a model to learn geographic location embeddings using Flickr tags, numerical environmental features, and categorical information. The experimental results show that our model can integrate Flickr tags with structured information in a more effective way than existing methods, leading to substantial improvements over baseline methods on various prediction tasks about the natural environment. 

\section*{Acknowledgments}
Shelan Jeawak has been sponsored by HCED Iraq. Steven Schockaert has been supported by ERC Starting Grant 637277. 

\bibliographystyle{splncs04}
\bibliography{ref.bib}

\end{document}